\documentclass[aps,a4paper,10pt,twocolumn,showpacs,longbibliography]{revtex4-1}

\usepackage[pdftex]{graphicx}	
	\DeclareGraphicsExtensions{.pdf,.png,.jpg}

\usepackage[applemac]{inputenc}
\usepackage{amsmath}
\usepackage{amssymb,amsthm}
\usepackage{mathbbol,bbm}
\usepackage{graphicx,float}
\usepackage[final]{showkeys}
\usepackage{bm,dsfont}
\usepackage{footnote}
\usepackage{lipsum}
\usepackage{color}

\def\ket#1{\left|#1\right>}
\def\bra#1{\left<#1\right|}

\begin{document}
\title{Iterative Phase Optimisation of Elementary Quantum Error Correcting Codes}

\author{M.~M\"uller$^{1}$, A.~Rivas$^{2}$, E.~A.~Mart\'inez$^{3}$, D.~Nigg$^{3}$, P.~Schindler$^{3}$, T.~Monz$^{3}$, R.~Blatt$^{3,4}$, and M.~A.~Martin-Delgado$^{2}$}
\vspace{2mm}
\affiliation{$^1$Department of Physics, Swansea University, Singleton Park, Swansea SA2 8PP, United Kingdom \\
$^2$Departamento de F\'isica Te\'orica I, Universidad Complutense, Avenida Complutense s/n, 28040 Madrid, Spain\\
$^3$Institut f\"ur Experimentalphysik, Universit\"at Innsbruck, Technikerstrasse 25, A--6020 Innsbruck, Austria \\
$^4$Institut f\"ur Quantenoptik und Quanteninformation, \"Osterreichische Akademie der Wissenschaften,Technikerstrasse 21A, 6020 Innsbruck, Austria}

\vspace{-3.5cm}

\date{\today}
\begin{abstract}
  Performing experiments on small-scale quantum computers is certainly
  a challenging endeavor. Many parameters need to be optimized to achieve high-fidelity
  operations. This can be done efficiently for operations acting on single qubits as errors can be fully
  characterized. For multi-qubit operations, though, this is no
  longer the case as in the most general case analyzing the effect of
  the operation on the system requires a full state tomography for which
  resources scale exponentially with the system
  size. Furthermore, in recent experiments additional electronic levels beyond the two-level system encoding the qubit
  have been used to enhance the capabilities of quantum information processors, which additionally increases the number of parameters that need to be controlled. For the optimization of the experimental system for a given task (e.g.~a quantum algorithm), one
  has to find a satisfactory error model and also efficient
  observables to estimate the parameters of the model. In this manuscript we
  demonstrate a method to optimize the encoding procedure for a small
  quantum error correction code in the presence of unknown but
  constant phase shifts. The method, which we implement here on a
  small-scale linear ion-trap quantum computer, is readily applicable
  to other AMO platforms for quantum information processing.
\end{abstract}

\maketitle

\section{Introduction}

The faithful execution of quantum algorithms, even on small-scale
prototype quantum computers, poses formidable control requirements
\cite{ladd-nature-464-45}. The influence of a multitude of error
sources and control parameters needs to be characterized and minimised
in order to enable overall high-fidelity operations. Within the field
of quantum control and optimisation, many techniques have been
developed
\cite{wiseman-book,viola-PRL,viola-science-293-2059,Lidar-PRA, Emerson-Science,Wallman-NJP-2015,Ticozzi-Review}
to characterize noise and decouple quantum systems to the highest possible degree from their
environment. This allows one to increase the fidelity of desired target
quantum operations under remaining, ultimately unavoidable, residual
sources of imperfections.

In particular, in a bottom-up approach to building quantum-information
hardware one usually optimises the performance of individual building
blocks such as e.g.~single- and two-qubit gate operations. In
principle, imperfections in few-qubit operations can be characterized
by full quantum process tomography.  However, it is much more
practical to use prior understanding of the dominant underlying noise
processes to design an efficient protocol to characterize, validate and finally reduce the resulting error sources.

In the following, we will separate the imperfections into
a non-reversible coupling to a larger environment
\cite{breuer-book,angel-book}, including fluctuations of control
parameters on the one hand, and unknown but constant unitary operations on the other hand. The latter
errors can, in principle, be compensated by measuring the
unknown operation and applying the inverse operation onto the system. A simple laboratory example are systematic single-qubit
phase shifts, which arise, e.g., if the frequency of the field driving the
qubit does not perfectly match the qubit transition frequency
\cite{haffner-physrep-469-155}. This transforms an initial state
$\alpha \ket{0} + \beta \ket{1}$ into $\alpha \ket{0} + \beta e^{i \phi}\ket{1}$ with unknown but constant
phase $\phi$. The phase shift $\phi$ can be measured systematically
with Ramsey-type experiments \cite{ramsey-pra-78-695,foot-book} and
furthermore compensated for by applying one single-qubit rotation
$U_\textrm{comp}=\exp (i \phi Z/2)$, where $Z$ denotes the third Pauli matrix \cite{nielsen-book}. This Ramsey-based phase detection and
compensation technique can be extended to certain classes of
multi-qubit states, such as e.g.~$n$-qubit GHZ states,
$\alpha \ket{0}^{\otimes n} + \beta e^{i \phi}\ket{1}^{\otimes n}$
\cite{nielsen-book}.

More general unitary errors can only be characterized by full quantum
state tomography which scales exponentially with the number of
qubits. Thus, it is highly desirable to design protocols that allow one to
efficiently and precisely determine specific systematic errors. An important 
class of such errors are unknown, though systematic, relative phases between 
the components of more complex quantum states. It should be noted, that the
propagation of single qubit phase shifts through complex algorithms
cannot be measured efficiently with generic methods that are algorithm
independent.

In this work we introduce and experimentally demonstrate a method that
allows one to compensate systematic, unknown, but constant phase shift
errors that arise in the encoding procedure of small quantum error
correcting codes \cite{Lidar-book}. We theoretically outline the
protocol, numerically study its performance and discuss how it was
successfully used in a recent experimental realisation of a 7-qubit
quantum error correcting code with trapped ions
\cite{nigg-science-345-302}. The iterative optimization protocol does
not rely on full quantum state tomography \cite{nielsen-book, Cramer-Nat-Comm-2010} and furthermore it is found
to converge very rapidly for small quantum error correcting codes. As
a consequence, the method can be experimentally applied ``in-situ'',
i.e.~it can be applied in real-time to optimize the experimental
performance. In fact, in the experiments of
Ref.~\cite{nigg-science-345-302}, the measurements and feedback steps
required by the algorithm to optimize the overall performance of the
whole encoding circuit were performed within a total time of a few
minutes. This is short compared to typical time scales on which
systematic parameter drifts take place
\cite{schindler-njp-15-123012}. Here, we apply the protocol to a case
where the encoding of logical states was achieved by a circuit of
unitary gate operations. However, similar scenarios where systematic,
constant phase shifts will arise in measurement-based encoding
protocols, can be addressed by the proposed technique
\cite{dennis-j-mat-phys-43-4452,Terhal-RevModPhys-2015}. Furthermore, the method is readily
applicable to other physical platforms for quantum information
processing, such as e.g.~Rydberg atoms \cite{jaksch-prl-85-2208,saffman-rmp-82-2313,Crow-arxiv-2015} in
optical lattices \cite{anderson-prl-107-263001,viteau-prl-107-060402,schauss-nature-491-87} or tweezer arrays
\cite{nogrette-prx-4-021034,xia-prl-114-100503}.

In the following two sections, we first briefly review some basic
properties of the implemented 7-qubit quantum error correcting code
\cite{steane-prl-77-793,bombin-prl-97-180501}, and then present in
some detail the experimental procedure used for the encoding of
logical quantum states. The latter discussion aims at illustrating
under which conditions the systematic phase shift errors which our
protocol tackles arise in the particular experiment of
Ref.~\cite{nigg-science-345-302}. Similar errors are expected to occur in
other atom- or solid-state based architectures \cite{hanson-nature-453-1043,Gambetta-Nat-Comm-2015,Gambetta-arxiv-2015,Kelly-nature-2015,Waldherr-Nature-2014,Fedorov-Nature-2012}, in particular those that exploit multi-level systems to enhance the systems' capability.

\begin{figure}[t]
\centering
\includegraphics[width=1\columnwidth]{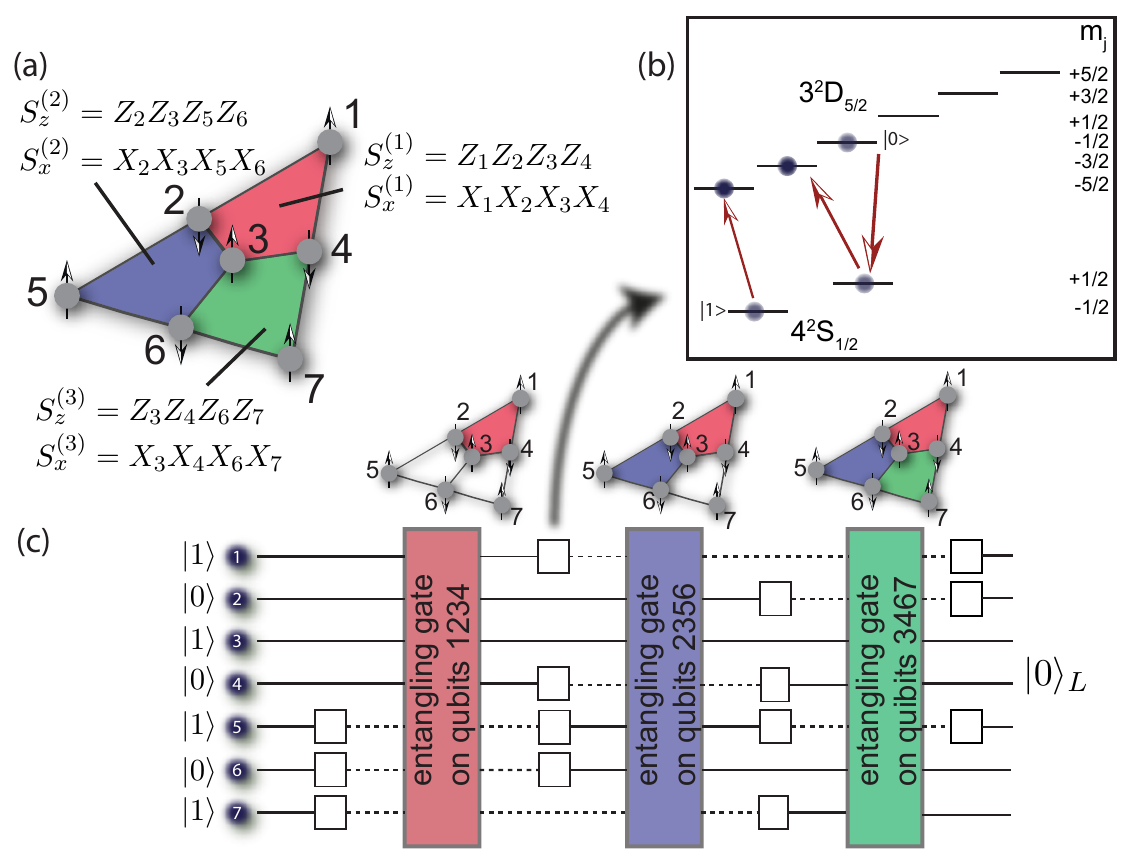}
\caption{\textbf{Schematics of the implemented 7-qubit quantum error
  correcting code and the encoding sequence.} (a) One
  logical qubit is encoded in seven physical qubits forming a
  two-dimensional triangular planar structure of three plaquettes. The
  code space is defined as the simultaneous +1 eigenspace of a set of
  six four-qubit stabilizer operators associated to the
  plaquettes. (b) Physical qubits are encoded in (meta-)stable
  electronic states of a string of seven $^{40}$Ca$^+$ ions. The
  computational subspace of each physical qubit is spanned by the two
  electronic states $4^{2}S_{1/2}(m_{j}=-1/2)$ ($\ket{1}$) and $3^{2}D_{5/2}(m_{j}=-1/2)$
  ($\ket{0}$). Another pair of states ($3^{2}D_{5/2}(m_{j}=-5/2)$ and $3^{2}D_{5/2}(m_{j}=-3/2)$) is
  used to spectroscopically decouple individual ion-qubits. Red arrows
  indicate sequences of pulses that are applied to realize this
  coherent decoupling (see Ref.~\cite{nigg-science-345-302} for more
  details). Decoupled ions (indicated by dashed lines in (c)), will
  ideally not participate in subsequent dynamics, until they are
  recoupled, i.e.~coherently transferred back into the computational
  subspace (solid lines in (c)). This technique enables the application
  of entangling gate operations, which are in this setup implemented by
  illuminating the entire ion string by a global laser beam
  \cite{schindler-njp-15-123012}, to subsets of four qubits belonging
  to a given plaquette. (c) The logical qubit is encoded by
  coherently mapping the product input state $\ket{1010101}$ onto the
  logical state $\ket{0}_L$ (see Eq.~(\ref{eq:ideal_state})). The
  quantum circuit combines spectroscopic decoupling and recoupling
  operations (white boxes) with plaquette-wise entangling operations
  that effectively create GHZ-type entanglement between qubits
  belonging to the same plaquette.}
\label{fig:color_code}
\end{figure}

\subsection{Ideal Encoding of a 7-Qubit Quantum Error Correcting Code}
In Ref.~\cite{nigg-science-345-302} a seven-qubit quantum error
correcting code has been demonstrated. This particular code
corresponds to the 7-qubit Steane code \cite{steane-prl-77-793} and
represents also the smallest instance of a 2D topological color code
\cite{bombin-prl-97-180501}. Since the realised quantum error
correcting code belongs to the class of CSS codes
\cite{calderbank-pra-54-1098, nielsen-book}, the code space is
generated as the simultaneous +1 eigenspace of a set of mutually
commuting stabiliser operators
$S^{(i)}_x$ and $S^{(i)}_z$ which are the product of Pauli $X$ and $Z$ operators, respectively, associated to subsets $\{i\}$ of qubits, see Fig.~\ref{fig:color_code}(a). Each generator is of $X$- or $Z$-type so that
$S^{(i)}_{x} \ket{\psi}_L = S^{(i)}_{z} \ket{\psi}_L = + \ket{\psi}_L $ holds for all subsets
$\{i\}$ and any encoded logical state $\ket{\psi}_L$. A 7-qubit code
with subsets as illustrated in Fig.~\ref{fig:color_code} represents
the minimal instance of a 2D color code. There, each plaquette
involves 4 physical qubits and hosts one 4-qubit $X$- and $Z$-type
stabiliser.

Encoding of a logical state $\ket{\psi}_L$ thus amounts to preparing
the system of physical qubits in the +1 eigenspace of all
stabilisers. The logical state $\ket{0}_L$, for instance, being a +1
eigenstate of the six plaquette generators as well as of the logical
$Z$-operator, $Z_L = \prod_{j=1}^{7} Z_j$, is explicitly
given by the following superposition of $2^3 = 8$ computational basis
states:

\begin{align}
\label{eq:ideal_state}
\ket{\psi_0} & =\frac{1}{2\sqrt{2}}(\ket{0000000}+\ket{0110110}+\ket{1111000} \nonumber \\
& +\ket{1001110}+\ket{0011011}+\ket{0101101} \nonumber \\
& + \ket{1100011}+\ket{1010101}).
\end{align}

\subsection{Experimental Encoding and Origin of Systematic Phase Shifts}

In Ref.~\cite{nigg-science-345-302} the outlined 7-qubit quantum error correcting code was
realised using a string of 7 trapped $^{40}$Ca$^+$ ions in a linear
Paul-trap based quantum computing architecture \cite{schindler-njp-15-123012}. Each
of the ions hosts one physical qubit encoded in the computational
subspace spanned by two (meta-)stable, electronic states, as shown in Fig.~\ref{fig:color_code}(b).

Arbitrary operations can be applied to the quantum register with the
following universal set of operations: Single qubit rotations can be
realised by a tightly focused laser beam illuminating single ions of
the string, whereas collective (non-entangling) rotations can be
implemented by a beam that collectively and homogeneously illuminates
the entire string of $n$ ions (see Ref.~\cite{schindler-njp-15-123012}). In addition,
a bichromatic laser field, illuminating the entire string of ions, is
used to implement a collective, $n$-qubit M{\o}lmer-S{\o}rensen (MS)
entangling gate operation \cite{molmer-prl-82-1835,roos-njp-10-013002}. Any arbitrary unitary operation
can be realized by a sequence of these operations that can be found
using refocusing techniques originally developed in NMR \cite{vandersypen-rmp-76-1037} or
numerical optimisation routines \cite{nebendahl-pra-79-012312}.

It is possible to extend the experimental toolbox by using more electronic levels than only the two electronic states of the qubit. This allows one to realize entangling operations on subsets of ions with less overhead than any known optimized sequence. Ions hosting physical qubits that are not supposed to participate in a given entangling operation, are coherently transferred to an additional set of
meta-stable electronic states which do not couple to the field that
generates the operations as shown in
Fig.~\ref{fig:color_code}(b). The quantum state of these decoupled
ions will ideally remain unaffected by the operation of the globally
applied, bi-chromatic laser field driving the qubit transition and
implementing the collective entangling MS gate
operation. Subsequently, decoupled ions can be re-coupled by
coherently mapping their state back into the qubit subspace.

This extended set of operations was used in Ref.~\cite{nigg-science-345-302} to realise the
encoding of an initial logical state, say $\ket{1}_L$, by a unitary
circuit: There, the 7-ion system was initially prepared in a product
state, say $\ket{1010101}$, thus being already a +1 eigenstate of the
set of three $Z$-type stabiliser operators. Preparation of the 7-qubit
system in the +1 eigenspace of the $X$-type stabilisers was then
realised by a sequence of three entangling operations, each acting on
subsets of four qubits belonging to the three plaquettes of the code,
respectively (see Fig.~\ref{fig:color_code}(c)). Each of the
effective 4-qubit MS gates creates GHZ-type entanglement between the
four qubits belonging to a given plaquette. The entangling gates were
interspersed by a series of on the order of hundred single-ion pulses
(see Ref.~\cite{nigg-science-345-302} and supplemental material therein for details) to spectroscopically decouple and
subsequently recouple ions that are supposed not to participate in the
action of a four-qubit plaquette-wise entangling operation.

Along the application of this encoding sequence, undesired systematic
phase shifts on all ions are generated and accumulate. These can be of various physical origins and unknown magnitude, arising e.g. from off-resonant light shifts on ions residing in the decoupled electronic states during the application of the MS gate operations. Note that in the present experiment these phase shifts do not vary significantly even over long data accumulation times of several minutes or longer, as the laser light causing these ac-Stark shifts is well stabilized to ensure proper operation of the entangling operations \cite{schindler-njp-15-123012}. Other possible origins of such shifts are differential magnetic shifts between the different electronic states used to define the computational subspace and the decoupling of qubits, and a detuning of the control fields from the qubit transition frequency due to a slowly varying laser frequency. Measuring and compensating for such a qubit detuning can be performed using techniques developed in the context of quantum metrology \cite{Q_METROLOGY}. It is important to note that the MS entangling gate operation
commutes with systematic phase shifts in the sense that the essential
part of the complex circuit, namely the three entangling gate
operations, still generate a final quantum state that is locally
equivalent to the ideal encoded state of Eq.~(\ref{eq:ideal_state}),
however with a set of unknown, relative phases $\{\phi_i\}$:
\begin{align}
\label{eq:non_ideal_state}
\ket{\psi_0'} & =\frac{1}{2\sqrt{2}}(\ket{0000000}+e^{i\phi_1}\ket{0110110}+e^{i\phi_2}\ket{1111000} \nonumber \\
& +e^{i\phi_3}\ket{1001110}+e^{i\phi_4}\ket{0011011}+e^{i\phi_5}\ket{0101101} \nonumber \\
& + e^{i\phi_6}\ket{1100011}+e^{i\phi_7}\ket{1010101}).
\end{align}
In order to maximize the fidelity of the encoded state these phases
need to be characterized and compensated for. There is no simple
Ramsey type experiment to determine these phases, hence we need to find a
protocol to measure them without full quantum state tomography.

\section{The Proposed Method}
\label{sec:method}

Some of the error sources in a quantum state preparation process, such as in the encoding discussed in the previous sections, result in ``true'' decoherence, which cannot be reversed by a subsequent application of unitary operations. The question whether a given source of imperfections results in systematic, coherent errors that can be calibrated out, or in decoherence, depends strongly on the origin of the noise and is related to the noise fluctuation time scale as compared to the data acquisition time. For instance, phase shifts that vary over the (short) times required to execute an individual or a few runs of a quantum circuit result in dephasing that the quantum error correcting procedure itself will take care of. In contrast, phase shifts that do not change their nature over (long) data acquisition times give rise to systematic coherent shifts that can be detected and compensated for. In the presented experiments, phase shifts are predominantly of this latter type as they are mainly caused by ac-Stark shifts originating from entangling operations that are performed on neighbouring qubits. These do not vary significantly over the data acquisition time required to implement the proposed phase optimization technique. The algorithm we propose aims at determining and undoing systematic unitary errors such as relative phase shifts in a simple, iterative manner without full state reconstruction. A simple model to outline the working principle of the proposed phase compensation technique is the formulation of the resulting final experimental state in the form of a Werner-type state,
\begin{equation}
\label{eq:Werner_state}
\rho = \frac{p}{\text{dim}} \mathds{1} + (1-p) \ket{\psi_0'} \bra{\psi_0'},
\end{equation}
where the part proportional to the identity operator, representing a completely mixed state, stands for a white-noise component, accounting for irreversible decoherence processes (dim = $2^7 = 128$ in the present case). The second term corresponds to the state $\ket{\psi_0'}$ (see Eq.~(\ref{eq:non_ideal_state})) containing a set of unknown phase shifts, which will be compensated by the application of corrective unitary phase shifts, in order to transform this component into the ideal encoded logical state $\ket{\psi_0}$ of Eq.~(\ref{eq:ideal_state}). The parameter $p \in [0, 1]$ quantifies the magnitude of the irreversible noise component, interpolating between the ideal target state (up to the unitary phase shifts) for $p=0$ and a fully mixed state in the limit $p=1$.

\begin{figure*}[t]
\centering
\includegraphics[width=0.8\textwidth]{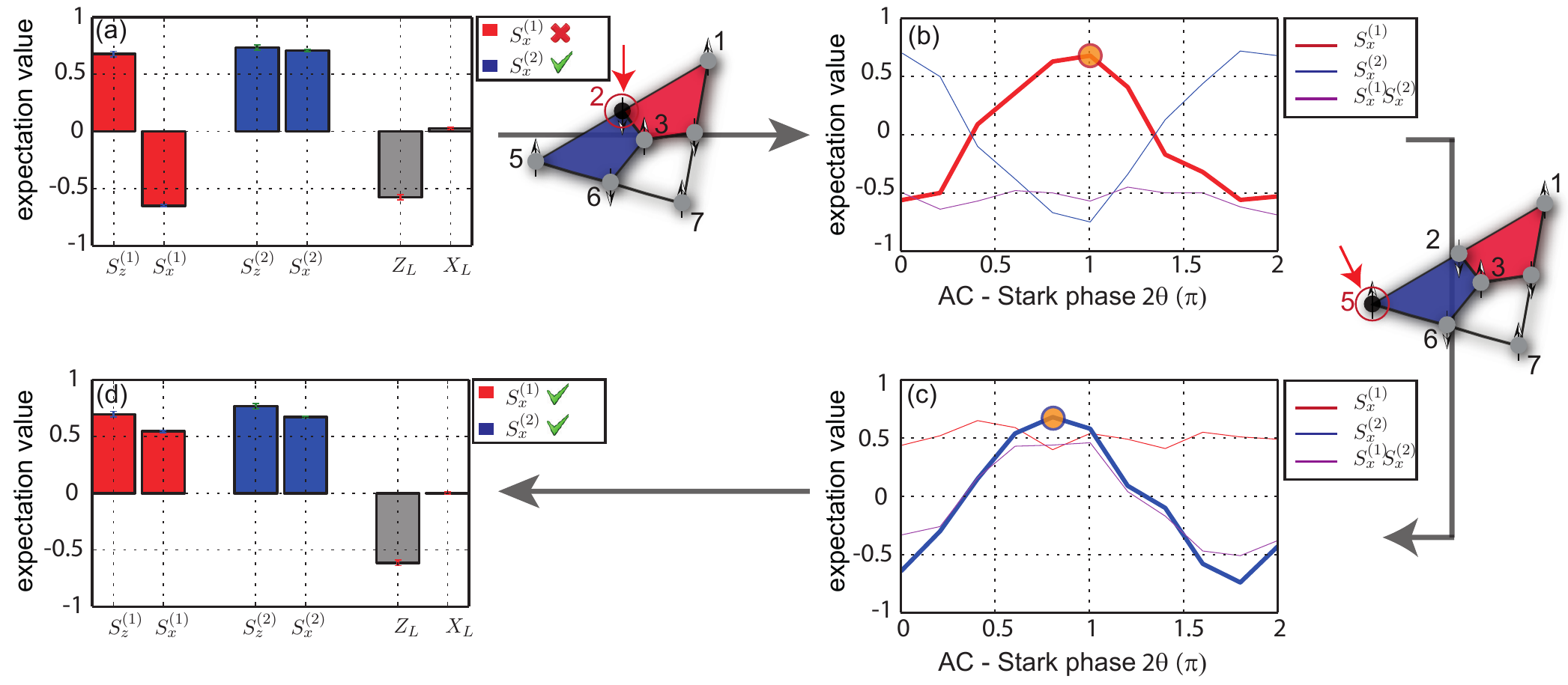}
\caption{\textbf{Experimental implementation of the phase optimisation protocol.} Here, the algorithm was applied to the intermediate state in the encoding sequence, which results from the application of the first two entangling operations acting on the qubits of the first (red) and second (blue) plaquette of the planar 7-qubit quantum error correcting code. The resulting state (a), before the application of the iterative phase optimisation technique, is characterised by positive values of $Z$-type plaquette stabilizer expectation values, which are maximal within the experimentally achieved accuracy of the encoding circuit~\cite{nigg-science-345-302}. On the other hand, $X$-type stabilizer expectation values have arbitrary values (positive on the first, negative on the second plaquette), indicating the presence of undesired, unknown relative phase shifts (see state Eq.~(\ref{eq:2plaquette_phase_shifted})). In the first step of phase optimisation (b), a $Z$-rotation of variable magnitude is applied to qubit \#2, which results in a sinusoidal behaviour of the expectation values of the stabizers $\langle S^{(1)}_x\rangle$ and $\langle S^{(2)}_x\rangle$ (cf.~Eqs.~(\ref{meanS1}) and (\ref{meanS2})), whereas the expectation value $\langle S^{(1)}_xS^{(2)}_x\rangle$ not containing $X_2$ remains constant. For each scan, the stabilizer that takes part in the optimization procedure, is highlighted by the bold line and the corresponding maximum value is marked via the orange circle. After reading off and fixing $\theta_2$ to the value which maximises $\langle S^{(1)}_x\rangle$ (orange circle), next a $Z$-rotation is applied to qubit \#5 (c). This scan is used to fix $\theta_5$ to the value which maximizes $\langle S^{(2)}_x\rangle$. Whereas in principle at this point one would proceed with the optimisation of $\langle S^{(1)}_xS^{(2)}_x\rangle$ by a $Z_1$-rotation scan, the data shows that all three stabilizers have within experimental resolution already reached the maximum, indicating convergence of the protocol. This is also reflected by both $X$-type plaquette stabilizers now being positive and maximal (d), while the expectation values of $Z$-type stabilizers and of the logical $Z$ operator have remained unchanged over the application of the algorithm -- compare (a) and (d). Experimental parameters: In each scan, different values for the phases characterising the single-qubit rotations were applied with an elementary step-size of $2\pi / 10$. For each phase value, the experiment was repeated 200 times.}
\label{fig:Phase_Step2}
\end{figure*}

For simplicity, we start by discussing the working principle of the phase compensation method for an intermediate state in the full encoding sequence shown in Fig.~\ref{fig:color_code}. This state we will optimize is the one that is reached after the application of the 4-qubit entangling operations to the first and the second plaquette of the planar, three-plaquette quantum error correcting code (see Fig.~\ref{fig:color_code}(a)). The ideal target state at this stage of the encoding sequence is given by
\begin{equation}
\label{eq:2plaquette_ideal}
\ket{\psi_0}=\frac{1}{2}(\ket{0000000}+\ket{0110110}+\ket{1111000}+\ket{1001110}).
\end{equation}
It maximizes the value of the generating $X$-type stabilizer operators on the first and second plaquette, $S^{(1)}_x=X_1X_2X_3X_4$ and $S^{(2)}_x=X_2X_3X_5X_6$, as well as of the stabilizer operator formed by the product of both, $S^{(1)}_xS^{(2)}_x$:
\begin{equation}
\langle \psi_0|S^{(1)}_x| \psi_0\rangle=\langle \psi_0|S^{(2)}_x| \psi_0\rangle=\langle \psi_0|S^{(1)}_xS^{(2)}_x| \psi_0\rangle=1.
\end{equation}
The state $\ket{\psi_0'}$ containing unknown phase shifts accumulated up to this point then reads
\begin{align}
\label{eq:2plaquette_phase_shifted}
\ket{\psi_0'}=&\frac{1}{2}(\ket{0000000}+e^{i\phi_1}\ket{0110110} \nonumber\\
&+e^{i\phi_2}\ket{1111000}+e^{i\phi_3}\ket{1001110}).
\end{align}
In order to compensate the relative phase shifts, we may apply single qubit $Z$-rotations to three of the six qubits, for instance
\begin{widetext}
\begin{equation}
  e^{i\theta_1 Z_1}e^{i\theta_2 Z_2}e^{i\theta_5 Z_5}\ket{\psi_0'} \rightarrow \frac{1}{2}(\ket{0000000}+e^{i[\phi_1+2(\theta_2+\theta_5)]}\ket{0110110}+e^{i[\phi_2+2(\theta_1+\theta_2)]}\ket{1111000}+e^{i[\phi_3+2(\theta_1+\theta_5)]}\ket{1001110}),
\end{equation}
\end{widetext}
where we have discounted the global phase factor $e^{-i(\theta_1+\theta_2+\theta_5)}$. The problem is to find the correct set of values $\bm{\theta}=[\theta_1,\theta_2,\theta_5]$ that compensates the phases and transform the state $\ket{\psi_0'}$ into $\ket{\psi_0}$. This can be viewed as an optimization problem as it is equivalent to finding the point $\bm{\theta}$ that is simultaneously a maximum of $\langle S^{(1)}_x\rangle$, $\langle S^{(2)}_x\rangle$ and $\langle S^{(1)}_xS^{(2)}_x\rangle$. Note that under the application of $Z$-type rotations, $Z$-type stabiliser expectation values remain unchanged.

Experimentally, an exhaustive search to determine the set of values of the three phases $\bm{\theta}$ which maximize the $X$-type stabilizers is impractical, as the number of possible phase configurations grows exponentially with the number of phases. Instead, we may apply the following \textbf{iterative protocol}:
\begin{enumerate}
\item \textbf{Fixing of the phase-to-stabiliser correspondence:} For each $X$-stabilizer, an associated control parameter $\theta_i$ which controls the compensation unitary $ \exp(i\theta_i Z_i)$, acting on ion $i$, is chosen. The particular assignment of stabilizer operators to phases $\bm{\theta}$ is somewhat arbitrary, however, it is important that a given $X$-stabiliser associated to a given phase depends on the application of the corresponding $Z_i$ rotation. This is the case if and only if the $X$-stabilizer under consideration contains the Pauli matrix $X_i$ corresponding to the ion $i$, and thus does not commute with a $Z_i$ rotation. Note that once a particular phase-to-stabiliser assignment is chosen, this should not be altered during subsequent steps of the optimization algorithm. Here, we choose $\theta_2$ for $S^{(1)}_x$, $\theta_5$ for $S^{(2)}_x$ and $\theta_1$ for $S^{(1)}_xS^{(2)}_x$, respectively.
  \item Choose an \textbf{initial configuration} for the set of rotation parameters $\bm{\theta}^{(0)}=[\theta_1^{(0)},\theta_2^{(0)},\theta_5^{(0)}]$.
  \item Experimentally \textbf{optimize} $\bm{S^{(1)}_x}$: The mean value of $S^{(1)}_x$ depends on control parameter $\theta_2$ in the following sinusoidal form,
  \begin{align}\label{meanS1}
    \langle S^{(1)}_x\rangle =&\frac{1}{2}\left\{\cos[\phi_2+2(\theta_1+\theta_2)]\right. \nonumber\\
    &+\left.\cos[\phi_1-\phi_3+2(\theta_2-\theta_1)]\right\}.
  \end{align}
  Scan $\theta_2$ over the interval $[0, 2\pi]$, while keeping $\theta_1 = \theta_1^{(0)}$ and $\theta_5 = \theta_5^{(0)}$ fixed. Measure all qubits in the $X$ basis to determine and fix $\theta_2$ to the value $\theta_2=\theta_2^{(1)}$ for which the measured mean value $\langle S^{(1)}_x\rangle$ is maximized.
  \item
  Next, experimentally \textbf{optimize} $\bm{S^{(2)}_x}$
  \begin{align}\label{meanS2}
    \langle S^{(2)}_x\rangle=&\frac{1}{2}\left\{\cos[\phi_1+2(\theta_2+\theta_5)]\right. \nonumber\\
    &+\left.\cos[\phi_2-\phi_3+2(\theta_2-\theta_5)]\right\},
  \end{align}
  by scanning $\theta_5$, while keeping the other control parameters at their previously determined values, i.e.~$\theta_2 = \theta_2^{(1)}$ and $\theta_1 = \theta_1^{(0)}$. Fix $\theta_5$ to the value $\theta_5=\theta_5^{(1)}$ which maximizes $\langle S^{(2)}_x\rangle$.
  \item Finally, apply a similar \textbf{optimization for} $\bm{S^{(1)}_xS^{(2)}_x}$,
  \begin{align}\label{meanS1S2}
    \langle S^{(1)}_xS^{(2)}_x\rangle=&\frac{1}{2}\left\{\cos[\phi_3+2(\theta_1+\theta_5)]\right.\nonumber \\
    &+\left.\cos[\phi_1-\phi_2+2(\theta_5-\theta_1)]\right\},
  \end{align}
  i.e.~scan over $\theta_1$ at fixed values $\theta_2=\theta_2^{(1)}$ and $\theta_5=\theta_5^{(1)}$, to find the value of $\theta_1 = \theta_1^{(1)}$ that maximizes $\langle S^{(1)}_xS^{(2)}_x\rangle$. This step completes one update round for the set of control parameters $\bm{\theta}=[\theta_1^{(0)},\theta_2^{(0)},\theta_5^{(0)}] \rightarrow [\theta_1^{(1)},\theta_2^{(1)},\theta_5^{(1)}]$.
  \item \textbf{Iterate until convergence is reached:} Repeat steps 3--5 $n$ times obtaining iteratively updated sets of values $\bm{\theta}^{(n)}=[\theta_1^{(n)},\theta_2^{(n)},\theta_5^{(n)}]$, until the set of phases $\bm{\theta}$ does -- within experimental resolution -- not change any further. For large enough values of $n$, the method converges to the maximal values of $\langle S^{(1)}_x\rangle$, $\langle S^{(2)}_x\rangle$ and $\langle S^{(1)}_xS^{(2)}_x\rangle$. Thereby, the component of the final state corresponding to $\ket{\psi_0'}$  of Eq.~(\ref{eq:2plaquette_phase_shifted}) is transformed, as desired, into the correct one $\ket{\psi_0}$ (see Eq.~(\ref{eq:2plaquette_ideal})).

If systematic phase shift errors were the only experimental source of imperfections, this maximal values would all be equal to one, corresponding to the case $p = 0$ in the model of Eq.~(\ref{eq:Werner_state}). In practice, decoherence processes are significant ($p > 0$) and reduce the experimentally attainable maximal values of the set of stabilizer operators.
\end{enumerate}

Figure \ref{fig:Phase_Step2} shows how the described phase optimisation algorithm works in experiment. Here, it was applied to remove relative phase shifts in the ideal, intermediate state Eq.~(\ref{eq:2plaquette_ideal}) after the first two entangling operations. Interestingly, the algorithm converges very quickly, namely already after performing two optimisation steps of stabilisers during the first round of iterations, $n=1$. Overall this resulted in a time of $\approx$7 minutes required for the application of the phase optimization protocol, as compared to about $\approx$48 minutes necessary for a full six-qubit state tomography under comparable conditions. Note that the required time for full state tomography does not include state reconstruction as well as phase optimization.

\section{Analysis and Properties of the Method}

As seen, the proposed phase optimization method provides correct results with very fast convergence for the two-plaquette case. Let us now analyze more in detail its mathematical background and performance for larger-dimensional optimization problems.
\subsection{Connection to coordinate descent/ascent methods}
To better explain the properties of the protocol and why it works, let us first consider a function of $\bm{\theta}=[\theta_1,\theta_2,\theta_5]$ defined as the sum of the stabilizer operators $\langle S^{(1)}_x\rangle$, $\langle S^{(2)}_x\rangle$ and $ \langle S^{(1)}_xS^{(2)}_x\rangle$ given in Eqs. \eqref{meanS1}, \eqref{meanS2} and \eqref{meanS1S2}:
\begin{equation}\label{f2}
  f(\bm{\theta}):=\langle S^{(1)}_x\rangle+\langle S^{(2)}_x\rangle+\langle S^{(1)}_xS^{(2)}_x\rangle.
\end{equation}
Instead of optimizing separately $\langle S^{(1)}_x\rangle$, $\langle S^{(2)}_x\rangle$ and $\langle S^{(1)}_xS^{(2)}_x\rangle$, we may maximize $f(\bm{\theta})$ following the same method as in the steps 3--5 above, i.e. fixing $\theta_2=\theta_2^{(0)}$ and $\theta_5=\theta_5^{(0)}$ and optimizing $f[\theta_1,\theta_2^{(0)},\theta_5^{(0)}]$ to obtain $\theta_1^{(1)}$; then repeating the procedure now fixing $\theta_1=\theta_1^{(1)}$ and $\theta_5=\theta_5^{(0)}$, and optimizing $f[\theta_1^{(1)},\theta_2,\theta_5^{(0)}]$ to obtain $\theta_2^{(1)}$, and so on.

This recipe is essentially a global version of coordinate descent (ascent) methods for minimizing (maximizing) functions of several variables, see \cite{wright-math-prog-151-3,nesterov-SIAM-22-341}. It is global in the sense that the optimization in every coordinate is done by searching the global maximum instead of applying gradient algorithms. It is clear from the very formulation of the method that $f$ will monotonically increase,
\begin{equation}
  f(\bm{\theta}^{(0)})\leq f(\bm{\theta}^{(1)}) \leq f(\bm{\theta}^{(2)}) \leq \ldots.
\end{equation}
Therefore, the only way that under this method $f$ might not converge to its maximum point is that it gets stuck in a local (but not global) maximum at some step. Nevertheless, one can show that the function $f(\bm{\theta})$ in Eq.~\eqref{f2} only has global maxima (see Appendix \ref{app:extrema_f}), so the recipe is guaranteed to work. In this regard, note that the method can work even for a function with local maxima as the optimization in every individual coordinate is done by seeking for the global maximum instead of applying differential methods which can present problems with local extremal points.

This argument regarding convergence of the method for $f(\bm{\theta})$ does not explain entirely the convergence when applied separately to $\langle S^{(1)}_x\rangle$, $\langle S^{(2)}_x\rangle$ and $ \langle S^{(1)}_xS^{(2)}_x\rangle$ as in steps 3--5 of the iterative algorithm outlined above. Nevertheless, the latter, experimentally used algorithm works as well because, on the one hand, the optimal point $\bm{\theta}$ for $\langle S^{(1)}_x\rangle$ is also optimal for $\langle S^{(2)}_x\rangle$ and $ \langle S^{(1)}_xS^{(2)}_x\rangle$. In other words, there exists a common optimum point for every term contributing to the sum in $f(\bm{\theta})$. On the other hand, despite the fact that the maximization process of some stabilizer will, in general, reduce the value of other stabilizers at intermediate steps, the global optimization in every coordinate rapidly overcomes this effect.\\

\subsection{Optimization of the entire 7-qubit encoding: the three plaquette case}
The practical applicability of the method has been tested and benchmarked by applying it to the more complex case of the entire encoding of the 7-qubit code. Here, the aim is to determine and remove the $2^3 - 1 = 7$ relative phases of the state Eq.~(\ref{eq:non_ideal_state}) in the preparation of the logical $\ket{0}_L$. The procedure works similarly as in the case of two plaquettes discussed above, however, here we need to apply $Z$-rotations to all seven qubits,
\begin{widetext}
\begin{align} \label{rotatedpsi}
\prod_{i=1}^7e^{i\theta_iZ_i}\ket{\psi_0'}\rightarrow\frac{1}{2\sqrt{2}}&(\ket{0000000}+e^{i[\phi_1+2(\theta_2+\theta_3+\theta_5+\theta_6)]}\ket{0110110}\nonumber\\
&+e^{i[\phi_2+2(\theta_1+\theta_2+\theta_3+\theta_4)]}\ket{1111000}+e^{i[\phi_3+2(\theta_1+\theta_4+\theta_5+\theta_6)]}\ket{1001110}\nonumber \\
&+e^{i[\phi_4+2(\theta_3+\theta_4+\theta_6+\theta_7)]}\ket{0011011}+e^{i[\phi_5+2(\theta_2+\theta_4+\theta_5+\theta_7)]}\ket{0101101}\nonumber\\
&+e^{i[\phi_6+2(\theta_1+\theta_2+\theta_6+\theta_7)]}\ket{1100011}+e^{i[\phi_7+2(\theta_1+\theta_3+\theta_5+\theta_7)]}\ket{1010101}),
\end{align}
\end{widetext}
to correct all phases by maximizing the seven expectation values of plaquette operators $\langle S^{(1)}_x\rangle$, $\langle S^{(2)}_x\rangle$, $\langle S^{(3)}_x\rangle$, $\langle S^{(1)}_xS^{(2)}_x\rangle$, $\langle S^{(2)}_xS^{(3)}_x\rangle$, $\langle S^{(1)}_xS^{(3)}_x\rangle$ and $\langle S^{(1)}_xS^{(2)}_xS^{(3)}_x\rangle$. The explicit expressions of these expectation values showing their dependences on the control parameters $\bm{\theta}=[\theta_1, \ldots, \theta_7]$ are given in the Appendix \ref{app:stabilizers}.

\begin{figure*}[t]
\centering
\includegraphics[width=0.95\textwidth]{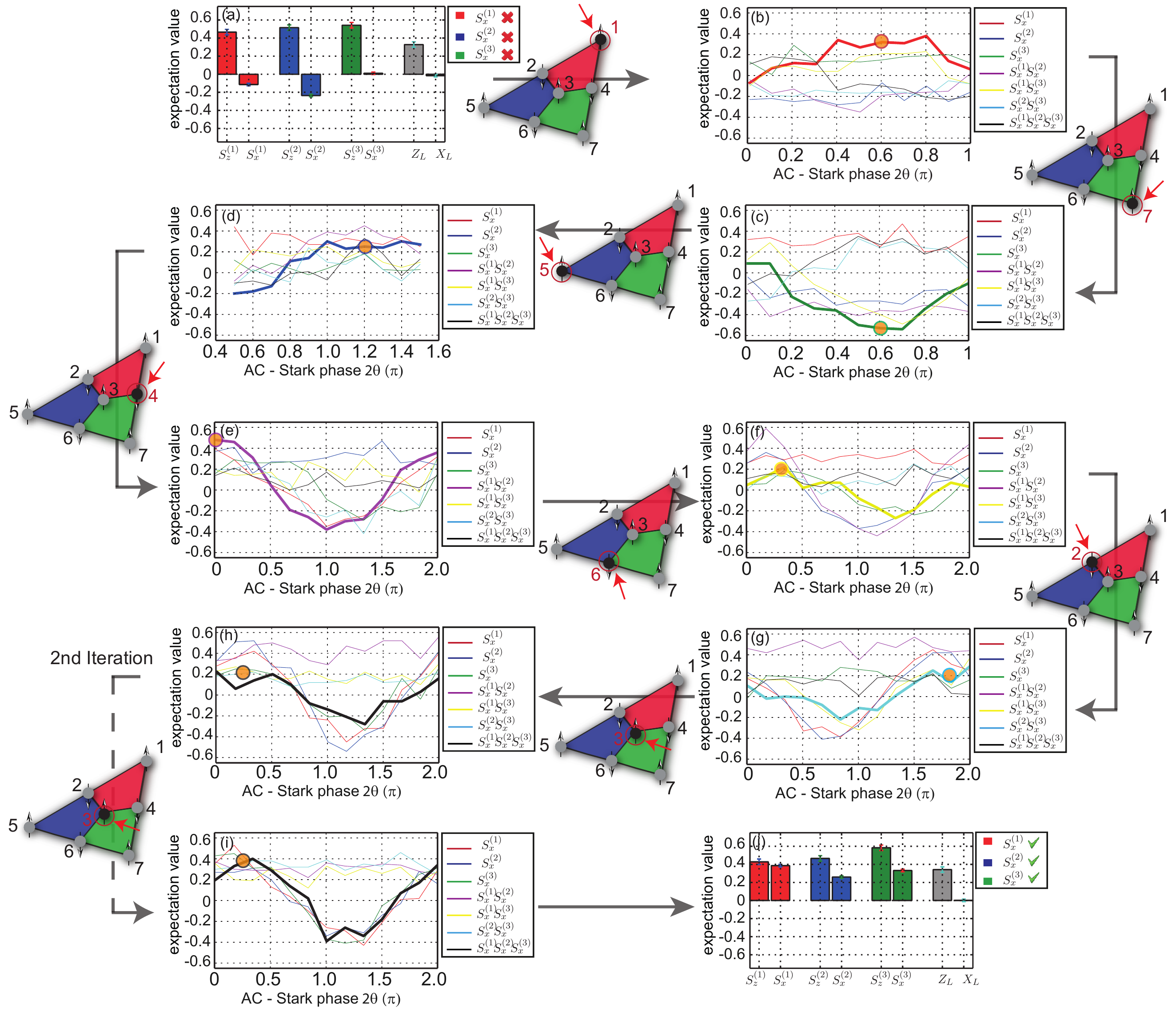}
\caption{\textbf{Experimental phase optimization of the complete 7-qubit quantum error correcting code.} Here, the algorithm was applied to the final state resulting from the complete encoding sequence shown in Fig.~\ref{fig:color_code}, i.e.~three entangling operations applied to the qubits belonging to the first (red), second (blue) and third (green) plaquette of the code. Initially, $X$-type stabiliser expectation values are non-maximal (a), indicating the presence of unknown, relative phases in the desired target state. After two rounds of iteratively maximising the seven expectation values of plaquette operators $\langle S^{(1)}_x\rangle$, $\langle S^{(2)}_x\rangle$, $\langle S^{(3)}_x\rangle$, $\langle S^{(1)}_xS^{(2)}_x\rangle$, $\langle S^{(2)}_xS^{(3)}_x\rangle$, $\langle S^{(1)}_xS^{(3)}_x\rangle$ and $\langle S^{(1)}_xS^{(2)}_xS^{(3)}_x\rangle$, the algorithm converges to a set of compensation phases, $\bm{\theta}=[\theta_1, \ldots, \theta_7]$, for which all $X$-type stabilisers assume maximal values. The individual phase value $\theta_i$ to the $Z$ rotation, which is adjusted to maximize the corresponding stabilizer expectation value under consideration (bold line), is indicated by the orange circle for each optimization step (see (b) - (i)). Note that due to the periodicity in $\theta_i$, it is also possible to search for the minimum expectation value of the stabilizer under consideration and adding the rotation angle $2\theta=\pi$, see (c) for an example. The $Z$ and $X$-type stabilisers of the logical state $|0\rangle_L$ after two rounds of optimization steps are shown in (j). Intermediate steps of the second round of optimisation are not shown. The experimental parameters are as specified in Fig.~\ref{fig:Phase_Step2}.}
\label{fig:Phase_Step3}
\end{figure*}

\subsection{Convergence on average}
Let us now obtain an estimate for the convergence rate. The function $f(\bm{\theta})$ for two plaquettes in Eq.~(\ref{f2}) can be written as a function of one component of the vector $\bm{\theta}$, say $\theta_1$, in the form of
\begin{equation}\label{ftheta1}
f(\theta_1)=A\cos(2\theta_1+\varphi)+c,
\end{equation}
where $A=A(\theta_2,\theta_5,\phi_1,\phi_2,\phi_3)$, $\varphi=\varphi(\theta_2,\theta_5,\phi_1,\phi_2,\phi_3)$ and $c=c(\theta_2,\theta_5,\phi_1,\phi_2,\phi_3)$. Specifically,
\begin{equation}
  A=\frac{1}{2}\sqrt{\cos^2\left[\theta_2+\theta_5+\frac{\phi_1}{2}\right] \cos ^2\left[ \theta_2-\theta_5+\frac{1}{2}(\phi_2-\phi_3)\right]}.
\end{equation}
By computing the mean value of this amplitude on a uniform distribution of their arguments we obtain $\bar{A}=0.81$ and similarly the mean value of $c$ is $\bar{c}=0$. In a rough, conservative estimate where the average value of $c$ remains constant when moving from the optimization of one coordinate of $\bm{\theta}$ to the next one, we estimate that in every coordinate optimization step we gain $\bar{A}/2=0.40$ on average. Therefore, since $\overline{f(\bm{\theta})}=0$ we estimate that we will obtain convergence after $n=2.47$ iterations on average. A similar estimate for the three plaquette case leads to a gain per coordinate optimization step of $\bar{A}/2=0.40$ on average, and thus estimated convergence after $n=2.47$ iterations on average, which is the same value as found for the 2-plaquette case.

These values can be checked by a numerical simulation of the method averaging over many random configurations of phases $\bm{\phi}$. For two and three plaquettes the simulation produces a mean value of $\bar{n}=1$ (exact) and $\bar{n}=2.25$ ($\bar{\sigma}=0.50$), respectively. On the one hand, the exact convergence after $n=1$ iterations for two plaquettes is due to the high degree of symmetry of $f(\bm{\theta})$ in that case, which has not been taken into account in the rough estimation of the average convergence rate. On the other hand, the simulation is compatible with the estimate for the three plaquette case, with a slightly improved, i.e.~faster rate of convergence.

To determine the convergence rate of the method optimizing individual mean values instead of their sum, we have numerically simulated this version of the algorithm used in the experiment, by averaging over random values of $\bm{\phi}$. This produces convergence after $\bar{n}=1$ (exact) iterations for the two plaquette case and $\bar{n}=2.16$ ($\bar{\sigma}=0.56$) for the three plaquette case.
The numerically observed convergence of $\bar{n}=1$ is in accordance with the experimentally observed convergence within a single iterative cycle of optimization (see discussion above and Fig.~\ref{fig:Phase_Step2}). Interestingly, the numerical results for the three plaquette case suggest that the variant based on optimizing individual mean values converges slightly faster than optimizing the sum over all of them.

Furthermore, as expected from the analytical arguments, our numerical study confirms did not encounter any phase configurations for which the optimization algorithm gets stuck or reaching convergence takes particularly long. In fact, the worst case in $10000$ random simulation runs corresponded to convergence after $n = 5$ iterative cycles. Further details can be found in Appendix \ref{app:concergence}.

\subsection{Experimental optimisation of the 7-qubit code}

Figure \ref{fig:Phase_Step3} shows experimental results of the iterative phase optimisation algorithm applied to the entire encoding sequence of the 7-qubit error correcting code. Whereas initially $X$-type stabilizer expectation values are non-maximal due to the presence of unknown relative phase shifts in the state of Eq.~(\ref{eq:non_ideal_state}), after two iterative cycles ($n=2$), composed of 14 elementary optimisation steps, the algorithm converges within the experimental resolution, and outputs a set of values for the compensation phase shifts $\bm{\theta}=[\theta_1, \ldots, \theta_7]$, for which the initially unknown relative phases $\{\phi_i\}$ are removed. As a consequence, not only the $Z$-type stabilizer values, which are unaffected by the optimisation protocol, but also all $X$-type stabilizers are positive-valued and maximal within the given accuracy of the encoding quantum circuit. The experimentally observed convergence after $n=2$ rounds is in very good agreement with the numerically prediction of $n=2.16$ for the three plaquette case.

\subsection{Scalability properties}

Let us now briefly discuss to which extent the present protocol is scalable as quantum states of systems of larger number of qubits are considered. In the analyzed two-plaquette case, we have been able to compensate three undesired relative phases by applying three qubit $Z$-rotations. For the complete minimal planar 7-qubit color code we needed to apply seven single-qubit $Z$-rotations. Larger instances of 2D color codes (see Fig.~5 in the Appendix \ref{app:counting}) encode logical qubits in a larger number of physical qubits and thereby provide larger logical distances and increased robustness to errors. The number of computational basis states involved as components in logical states of such larger systems grows exponentially with the number of plaquettes \#, and so does the number \#of relative phases that need to be compensated:
\begin{equation}
\#\text{Phases}=2^{\#\text{Plaquettes}}-1.
\end{equation}
In the most general case, these relative phases may be uncorrelated among each other, so that an exponential number of independent $Z$-type Hamiltonian generators are required to unitarily compensate all phases. This can in principle be achieved by resorting not only to single-qubit $Z$-rotations, but also to two-qubit $ZZ$-rotations, $\exp(\theta_{ij} Z_i Z_j)$, three-qubit, and higher-order $n$-body rotations. Following this route the required operations become more and more nonlocal. One can then ask to which size of a planar color code the method can be extended such that only physically quasi-local rotations, i.e.~$n$-qubit rotations only acting on qubits belonging to the same plaquette, are sufficient to correct the set of undesired phases. Combinatorics show (see Appendix \ref{app:counting}) that phases in the state of a logical distance $d=5$ color code involving 17 qubits can in principle still be corrected by such physically quasi-local rotations, whereas the next-larger generation, a distance $d=7$ color code encoded in 31 qubits would require physically non-local rotations acting on qubits on several plaquettes.

This mismatch between degrees of freedom and local operations, which becomes more significant as the code size increases, is a generic feature and not specific to color codes. It will ultimately need to be circumvented by the implementation of quantum error correcting codes in physical architectures where physical error sources act quasi-locally, and by using fault-tolerant encoding protocols \cite{shor96a, preskill-review}, which avoid an uncontrolled propagation of errors during the encoding over the entire quantum hardware \cite{dennis-j-mat-phys-43-4452, Terhal-RevModPhys-2015}.

\section{Conclusions and Outlook}
In this work we have proposed and experimentally shown an iterative phase optimisation protocol that allows one to efficiently compensate systematic, unknown but constant phase shift errors, which can occur e.g.~in realizations of small quantum error correcting codes. The method allows one to determine and remove such relative phases without full quantum state tomography, and it converges very quickly when applied to small quantum error correcting codes. This algorithm was a key element in optimizing a recent successful implementation of a 7-qubit quantum error correcting code in a system of trapped ions \cite{nigg-science-345-302}. The method can be equally applied to alternative, non-unitary encoding protocols based e.g.~on Quantum non-demolition (QND) measurements of stabilizer operators. Furthermore, the protocol demonstrated here is not limited to trapped ion-systems, and we hope that it will be useful also for other, currently ongoing efforts in quantum computing and error correction in AMO and solid-state systems.

\section*{Acknowledgments}
M.M. thanks M. Guta for valuable discussions. We gratefully acknowledge support by the Austrian Science Fund (FWF), through the SFB FoQuS (FWF Project No. F4002-N16), as well as the Institut f\"ur Quantenoptik und Quanteninformation GmbH. E.A.M. is a recipient of a DOC fellowship from the Austrian Academy of Sciences. P.S. was supported by the Austrian Science Foundation (FWF) Erwin Schr\"odinger Stipendium 3600-N27. The research is based upon work supported by the Office of the Director of National Intelligence (ODNI), Intelligence Advanced Research Projects Activity (IARPA), via the U.S. Army Research Office Grant No.W911NF-16-1-0070. The views and conclusions contained herein are those of the authors and should not be interpreted as necessarily representing the official policies or endorsements, either expressed or implied, of the ODNI, IARPA, or the U.S. Government. The U.S. Government is authorized to reproduce and distribute reprints for Governmental purposes notwithstanding any copyright annotation thereon. Any opinions, findings, and conclusions or recommendations expressed in this material are those of the author(s) and do not necessarily reflect the view of the U.S. Army Research Office. We also acknowledge support by U.S. A.R.O. through Grant No. W911NF-14-1-010, the Spanish MINECO Grant No. FIS2012-33152, and the CAM Research Consortium QUITEMAD+ S2013/ICE-2801.

\appendix

\begin{widetext}

\section{Phase dependences of stabilizer operators}
\label{app:stabilizers}
The expectation values of the seven stabilizer plaquette operators for the state Eq.~\eqref{rotatedpsi} are given by
\begin{align}
  \langle S^{(1)}_x\rangle=\tfrac{1}{4}&\{\cos[\phi_2+2(\theta_1+\theta_2+\theta_3+\theta_4)]+\cos[\phi_1-\phi_3+2(-\theta_1+\theta_2+\theta_3-\theta_4)]\nonumber\\
  &+\cos[\phi_4-\phi_6+2(-\theta_1-\theta_2+\theta_3+\theta_4)]+\cos[\phi_5-\phi_7+2(-\theta_1+\theta_2-\theta_3+\theta_4)]\},
\end{align}
\begin{align}
  \langle S^{(2)}_x\rangle=\tfrac{1}{4}&\{\cos[\phi_1+2(\theta_2+\theta_3+\theta_5+\theta_6)]+\cos[\phi_2-\phi_3+2(\theta_2+\theta_3-\theta_5-\theta_6)]\nonumber\\
  &+\cos[\phi_4-\phi_5+2(-\theta_2+\theta_3-\theta_5+\theta_6)]+\cos[\phi_6-\phi_7+2(\theta_2-\theta_3-\theta_5+\theta_6)]\},
\end{align}
\begin{align}
  \langle S^{(3)}_x\rangle=\tfrac{1}{4}&\{\cos[\phi_4+2(\theta_3+\theta_4+\theta_6+\theta_7)]+\cos[\phi_1-\phi_5+2(\theta_3-\theta_4+\theta_6-\theta_7)]\nonumber\\
  &+\cos[\phi_2-\phi_6+2(\theta_3+\theta_4-\theta_6-\theta_7)]+\cos[\phi_3-\phi_7+2(-\theta_3+\theta_4+\theta_6-\theta_7)]\},
\end{align}
\begin{align}
  \langle S^{(1)}_xS^{(2)}_x\rangle=\tfrac{1}{4}&\{\cos[\phi_3+2(\theta_1+\theta_4+\theta_5+\theta_6)]+\cos[\phi_1-\phi_2+2(-\theta_1-\theta_4+\theta_5+\theta_6)]\nonumber\\
  &+\cos[\phi_4-\phi_7+2(-\theta_1+\theta_4-\theta_5+\theta_6)]+\cos[\phi_5-\phi_6+2(-\theta_1+\theta_4+\theta_5-\theta_6)]\},
\end{align}
\begin{align}
  \langle S^{(1)}_xS^{(3)}_x\rangle=\tfrac{1}{4}&\{\cos[\phi_6+2(\theta_1+\theta_2+\theta_6+\theta_7)]+\cos[\phi_1-\phi_7+2(-\theta_1+\theta_2+\theta_6-\theta_7)]\nonumber\\
  &+\cos[\phi_2-\phi_4+2(\theta_1+\theta_2-\theta_6-\theta_7)]+\cos[\phi_3-\phi_5+2(\theta_1-\theta_2+\theta_6-\theta_7)]\},
\end{align}
\begin{align}
  \langle S^{(2)}_xS^{(3)}_x\rangle=\tfrac{1}{4}&\{\cos[\phi_5+2(\theta_2+\theta_4+\theta_5+\theta_7)]+\cos[\phi_1-\phi_4+2(\theta_2-\theta_4+\theta_5-\theta_7)]\nonumber\\
  &+\cos[\phi_2-\phi_7+2(\theta_2+\theta_4-\theta_5-\theta_7)]+\cos[\phi_3-\phi_6+2(-\theta_2+\theta_4+\theta_5-\theta_7)]\},
\end{align}
\begin{align}
  \langle S^{(1)}_xS^{(2)}_xS^{(3)}_x\rangle=\tfrac{1}{4}&\{\cos[\phi_7+2(\theta_1+\theta_3+\theta_5+\theta_7)]+\cos[\phi_1-\phi_6+2(-\theta_1+\theta_3+\theta_5-\theta_7)]\nonumber\\
  &+\cos[\phi_2-\phi_5+2(\theta_1+\theta_3-\theta_5-\theta_7)]+\cos[\phi_3-\phi_4+2(\theta_1-\theta_3+\theta_5-\theta_7)]\}.
\end{align}

\section{Extrema of the function $f(\bm{\theta})$}
\label{app:extrema_f}
By considering all variables of $\bm{\theta}$ fixed except one, say $\theta_1$, we have seen that the function $f(\bm{\theta})$ can be written as a cosine, Eq. \eqref{ftheta1}. Since this is true for every variable of $\bm{\theta}$ when fixing the rest of them, the sections of the function $f(\bm{\theta})$ in every variable are just cosine functions (one frequency). In such a situation it seems not to be possible to obtain local maxima or minima. This is because the hypersurface $f(\bm{\theta})$ can be viewed as a modulation of a cosine profile along all the orthogonal directions by other cosine profiles; since the cosines do not have local extrema, their modulations do not create local extrema. In fact the maximum (minimum) points are just the points which maximize (minimize) all sections individually. This can be checked with the two-plaquette case where the condition for the critical point of $f(\bm{\theta})$ is
\begin{equation}
\nabla f(\bm{\theta})=-4\begin{bmatrix}
 \cos \left(\theta_2+\theta_5+\frac{\phi_1}{2}\right) \cos \left( \theta_2-\theta_5+\frac{\phi_2}{2}-\frac{\phi_3}{2}\right) \sin \left( 2\theta_1-\frac{\phi_1}{2}+\frac{\phi_2}{2}+\frac{\phi_3}{2}\right)\\
 \cos \left(\theta_1+\theta_5+\frac{\phi_3}{2}\right) \cos \left( \theta_1-\theta_5-\frac{\phi_1}{2}+\frac{\phi_2}{2}\right) \sin \left( 2\theta_2+\frac{\phi_1}{2}+\frac{\phi_2}{2}-\frac{\phi_3}{2}\right)\\
 \cos \left(\theta_1+\theta_2+\frac{\phi_2}{2}\right) \cos \left( \theta_1-\theta_2-\frac{\phi_1}{2}+\frac{\phi_3}{2}\right) \sin \left( 2\theta_5+\frac{\phi_1}{2}-\frac{\phi_2}{2}+\frac{\phi_3}{2}\right)
\end{bmatrix}=\bm{0}.
\end{equation}
By solving the three simultaneous conditions we obtain that the critical points which are a maximum or a minimum (the rest are saddle points) are
\begin{equation}
\bm{\theta}_c=\left(\frac{\phi_1}{4}-\frac{\phi_2}{4}-\frac{\phi_3}{4},-\frac{\phi_1}{4}-\frac{\phi_2}{4}+\frac{\phi_3}{4},-\frac{\phi_1}{4}+\frac{\phi_2}{4}-\frac{\phi_3}{4}\right)+\frac{\pi}{2}(k_1,k_2,k_3)
\end{equation}
where $k_1,k_2,k_3\in\mathds{Z}$. However, in this case
\begin{equation}
  f(\bm{\theta}_c)=(-1)^{(k_1 + k_2)} + (-1)^{(k_1 + k_3)} + (-1)^{(k_2 + k_3)}
\end{equation}
reaches either its absolute maximum (3) or its absolute minimum ($-1$).

\end{widetext}

\section{Convergence scaling}
\label{app:concergence}

\begin{figure}[t]
\centering
\includegraphics[width=1\columnwidth]{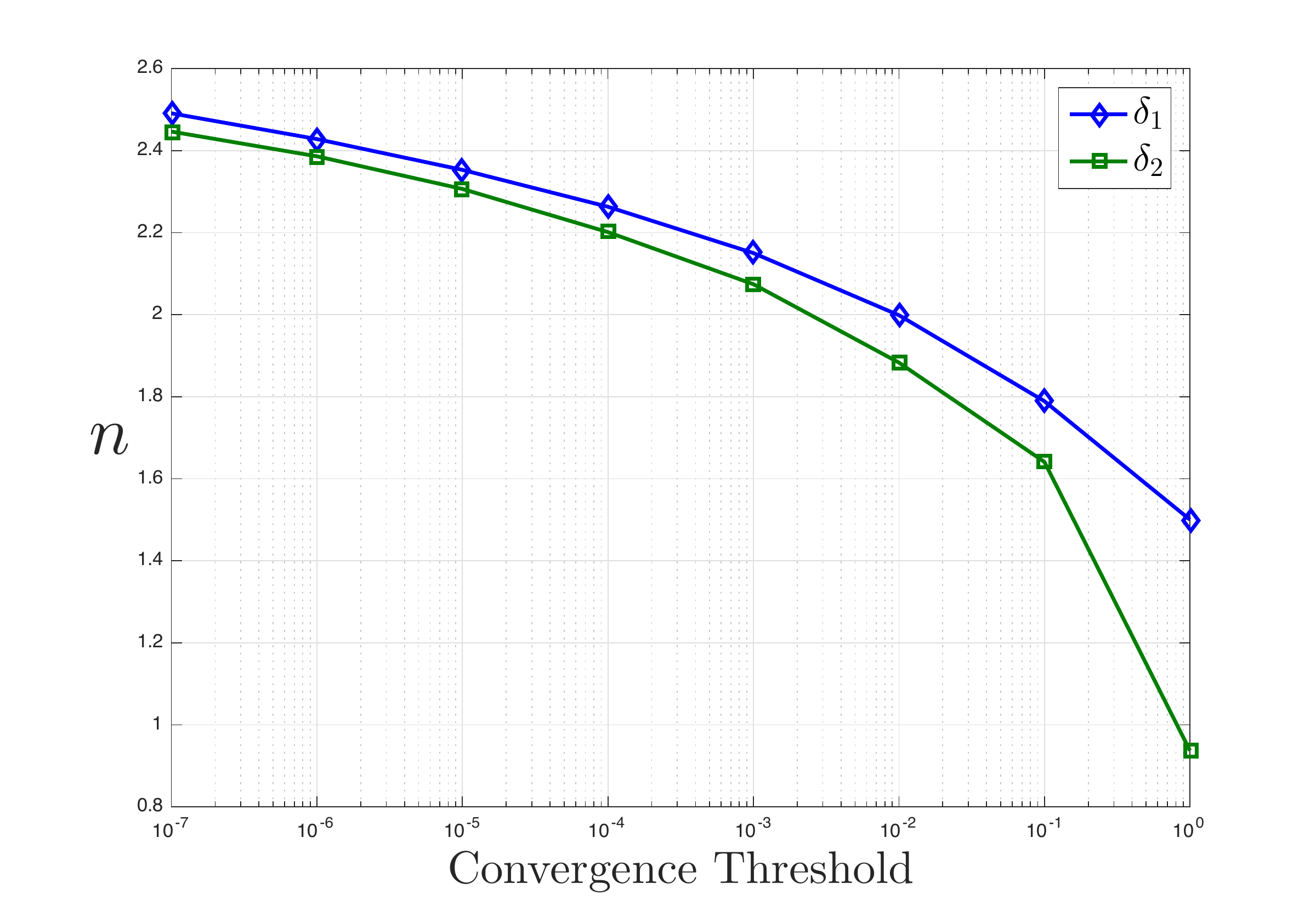}
\label{fig:convergence}
\caption{\textbf{Number of iterations vs. convergence threshold.}  This plot shows the scaling of the number of iterations required by PHOM with the tightness of the convergence criterion. As commented in the text, two figures of merit assess this, $\delta_1$ and $\delta_2$. The former is related to the distance between the sum of stabilizers and its maximum value, and the latter is associated with the maximum value among the distances for each stabilizer. The simulations have been done for the case of PHOM applied to individual mean values.}
\end{figure}

For practical purpuses, the average number of iterations required by the phase optimization method (PHOM) depends on the value taken as a convergence threshold, or equivalently on how close we demand the stabilizer mean values to approach their maximum value. In our case, we establish that convergence of the iterative optimization is reached once all stabiliser expectation values have assumed their maximal values to within $10^{-3}$. This is well within the experimental measurement accuracy \cite{schindler-njp-15-123012,nigg-science-345-302}, for which convergence is reached in practice.

For the sake of completeness, in Fig.~4 we show the average number of iterations as a function of the convergence threshold. We have quantified the latter by means of two figures of merit, namely $\delta_1=|f(\bm{\theta})-7|$ and $\delta_2$ that corresponds to the maximum among the distances of each individual stabilizer and its maximum value. Notably, fast convergence is observed throughout the whole range of numerical values considered.

\section{Number of quasi-local control degrees of freedom}
\label{app:counting}

\begin{figure}[t]
\centering
\includegraphics[width=1\columnwidth]{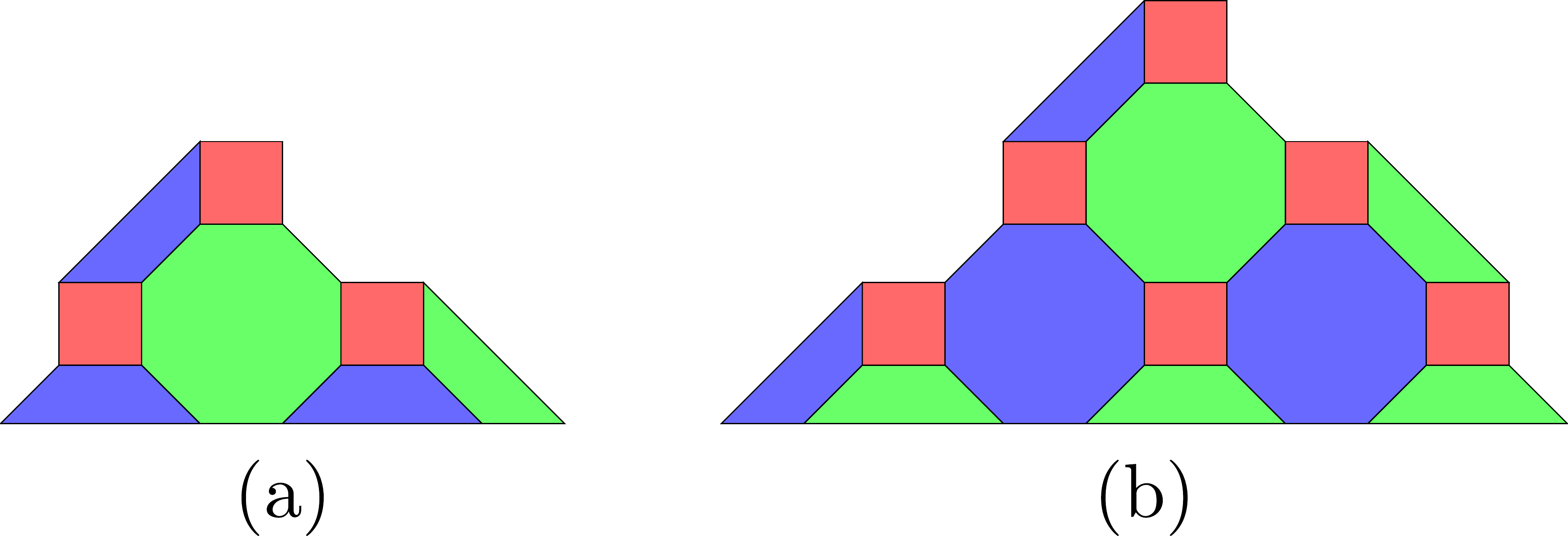}
\label{fig:23generation}
\caption{\textbf{Larger instances of planar color codes.}  The 17-qubit code (a) encodes a logical qubit of logical distance $d=5$, the 31-qubit code (b) has distance $d=7$. Whereas the 17-qubit code would, at least in principle, allow for the correction of the undesired phases with physically quasi-local rotations, acting only on subsets of qubits belonging to the same plaquette, phase compensation for the 31 qubit case (and larger codes) would require non-local rotations involving qubits of several plaquettes.}
\end{figure}

For the code with 17 qubits (distance $d=5$) we have 8 plaquettes (see Fig.~5), so the number of undesired relative phases is $2^{8}-1=255$. Counting the number of degrees of freedom we have available with plaquette Z-rotations yields the following numbers of $n$-local operations (i.e~operations involving $n$ qubits):
\begin{description}
\item[1-local] 17 one-qubit rotations.
\item[2-local] There are 7 square plaquettes which share 6 sides, and 1 octagonal plaquette which shares 6 sides with square plaquettes, so that, square plaquettes: $7\times \binom{4}{2}-6=36$, octagonal plaquette: $\binom{8}{2}-6=22$, Total: 58.\\
\item[3-local] Square plaquettes: $7\times \binom{4}{3}=28$,
octagonal plaquette: $\binom{8}{3}=56$.\\
\item[4-local] Square plaquettes: $7\times \binom{4}{4}=7$,
octagonal plaquette: $\binom{8}{4}=70$.\\
\item[5-local] Octagonal plaquette: $\binom{8}{5}=56$.\\
\end{description}
Therefore, taking into account up to 5-local rotations we obtain 292 degrees of freedoms. Thus, indeed only local plaquette rotations are sufficient to correct the undesired 255 phases in this 2nd generation of color codes.

The next code in the family, the one with 31 qubits (3rd generation, distance $d=7$), has 15 plaquettes (Fig.~5) and so it requires $2^{15}-1=32767$ rotations. A similar counting as in the 17 qubit case shows that the number of phases that can be corrected by quasi-local rotations only involving qubits belonging to the same plaquette is 875. Henceforth, it requires physically non-local rotations involving qubits of several plaquettes.

\bibliography{refs}

\end{document}